\documentclass[11pt]{cernrep}
\usepackage{latexsym,amssymb,amsfonts,mathbbol,color,epsfig,graphicx,
amsmath,mathrsfs,axodraw}
\newcommand{\be}{\begin{equation}}
\newcommand{\ee}{\end{equation}}
\newcommand{\bea}{\begin{eqnarray}}
\newcommand{\eea}{\end{eqnarray}} 
\newcommand{\bml}{\begin{mathletters} \baselineskip 10pt}
\newcommand{\eml}{\baselineskip 12pt \end{mathletters}}

\newcommand{\m}{{\scriptscriptstyle -}}
\newcommand{\p}{{\scriptscriptstyle +}}

\newcommand{\intl}{\int\limits_{-L}^L}

\newcommand{\bra}{\langle}
\newcommand{\ket}{\rangle}

\newcommand{\sfrac}[2]{{\textstyle \frac{#1}{#2}}}
\newcommand{\pad}[2]{\frac{\partial #1}{\partial #2}}

\newcommand{\prp}[1]{\vc{#1}_{\!\scriptstyle\perp}}

\newcommand{\ssprp}[1]{\ssvc{#1}_{\!\scriptscriptstyle\perp}}
\newcommand{\vc}[1]{\mbox{\boldmath$#1$}}

\newcommand{\ssvc}[1]{\mbox{\scriptsize\boldmath$#1$}}

\newcommand{\Tr}{\mbox{Tr}}

\begin{document}
\title{LIGHT-CONE ZERO MODES REVISITED}

\author{T.~Heinzl}
\institute{Theoretisch-Physikalisches Institut,
Friedrich-Schiller-Universit\"at Jena, 
Germany}
%

\maketitle
\begin{abstract}
The vacuum problem of light-cone quantum field theory is reanalysed
from a functional-integral point of view. 
\end{abstract}

\section{INTRODUCTION}

It has become a standard lore that ``the light-cone (LC) vacuum is
trivial''\footnote{See e.g.~the reviews \cite{revs} for extensive discussions
of this issue.}. This statement, however, appears somewhat paradoxical,
as many important physical phenomena are usually attributed to the existence
of a \textit{non}trivial vacuum. An incomplete list includes the notions of
condensates, spontaneous symmetry breaking, vacuum tunneling
(instantons) and decay, the Casimir energy and the cosmological
constant. In order to clarify whether these features can be reconciled
with a trivial vacuum, a necessary first step consists in precisely
clarifying what is meant by the concept of a trivial vacuum.     
In this contribution we denote a vacuum as trivial if the
associated vacuum persistence amplitude (VPA) is unity,
\be
\label{VPA1}
  1 = \bra 0 |\hat{S} | 0 \ket \equiv 
  \exp(iW [0]) \; . 
\ee
This is the probability amplitude that the vacuum state `persists' in
time, i.e. that the vacuum $|0 \ket$ in the remote past evolves to the
vacuum in the distant future under the influence of the interaction
encoded in the S--matrix $\hat{S}$. For a stable vacuum (no vacuum decay),
this is a phase, $\exp (iW[0])$, and triviality of the vacuum means that
the Schwinger functional vanishes, $W[0]=0$. Note that we do not make a
statement about the vacuum \textit{state} $|0\ket$ itself. Graphically,
(\ref{VPA1}) can be represented as 
\be
  \hspace{.5cm} 1 = \exp  \left\{ \rule[-.9cm]{0pt}{1.5cm}
  \CArc(20,3)(15,0,360) \CArc(50,3)(15,0,360) \Text(80,3)[]{$+$}
  \CArc(110,3)(15,0,360) \CArc(140,3)(15,0,360) \CArc(170,3)(15,0,360)
  \Text(200,3)[]{$+$}   \Oval(245,3)(15,30)(0)
  \CArc(245,3)(30,0,360) \Text(305,3)[]{$+ \quad \ldots$}
  \hspace{12cm}\right\} \; , 
\ee
so that $W[0] = 0$ is the statement that all vacuuum bubbles
vanish. This is totally consistent with the absence of
terms consisting of only creators  or annihilators ($a^\dagger
a^\dagger a^\dagger a^\dagger$, $aaaa$) in the LC Hamiltonian, which in turn
implies that the interaction does not change the vacuum,
\be
\label{LCHAM}
  H_{0} \,  | 0 \ket = 0 = H_{\mathrm{int}} \, | 0 \ket \; .
\ee
Writing the S--matrix in terms of the standard Dyson series, $\hat{S}
= T^\p \exp (-i \int dx^\p \, H_{\mathrm{int}})$, we are led back to
(\ref{VPA1}). On the other hand, the VPA is alternatively given by the
path integral  
\be
\label{VPA2}
  1 \stackrel{?}{=} \int \mathscr{D} \phi \, \exp i S[\phi] \equiv Z[0]
  \equiv \exp i W[0] \; ,
\ee
where $S$ denotes the classical action. Now, according to Feynman, the
path integral constitutes a `space--time approach to quantum theory' or,
put differently, it is explicitly frame independent. The question mark
in (\ref{VPA2}) represents the fact that in general the path integral is
different from unity, $Z[0] \ne 1$, and this is an invariant statement
which does not depend on the choice of coordinates (LC or else).
This apparent paradox is easily resolved by recalling that the VPA or
path integral is a mere normalization constant: in the generating
functional $T[J]$ of Green functions it is just divided out leading to
\be
  T[0] \equiv \left. Z[J]/Z[0] \right|_{J=0} = 1 \; ,
\ee
which \textit{is} unity by construction. The transition from $Z$ to
$T$ can be viewed as a renormalization of the cosmological constant or,
more practically, as the prescription to `omit the vacuum
bubbles'. Using the exponentiation (\ref{VPA2}) of vacuum bubbles we may
write 
\be
\label{T}
  T[J] = \int \mathscr{D} \phi \, \exp i \left\{ S[\phi , J] - W[0]
  \right\} \;,
\ee
from which we infer that the modified action $S' \equiv S - W[0]$ has
a trivial vacuum by construction. This suggests that the `standard' LC
Hamiltonian (\ref{LCHAM}) should correspond to the action $S'$ rather
than $S$. Of course, $W[0]$ in general is a divergent quantity so that,
prior to renormalization, (\ref{T}) is a somewhat formal statement.  
In what follows we will try to shed some light on the
mechanism by which the subtraction of vacuum bubbles arises.

\section{THE LC PATH INTEGRAL}

While the path integral itself does not depend on the choice of
coordinates, the integration measure as well as particular choices of
boundary conditions and sources do\footnote{See e.g.~the nice
discussion of the Casimir effect in \cite{lenz}}. In this sense we may
talk about a LC path integral. Consider now $\phi^4$-theory in $d$
dimensions. The VPA in the presence of a source $J$ is given by the
functional integral 
\be
  Z[J] = \int \mathscr{D} \phi \, \exp i \left\{ S_0 [\phi] +
  \frac{\lambda}{4!} (\phi^2, \phi^2) + (\phi, J) \right\} 
\ee
with the free action $S_0$ being the `quadratic form' $S_0 [\phi] \equiv
- \sfrac{1}{2} \, (\phi, \mathsf{K}_m \, \phi)$, defined in terms of the
Klein-Gordon operator $\mathsf{K}_m \equiv \Box + m^2$, the inverse of
which is the Feynman propagator $\Delta_m$. It is instructive to have a
look at its mixed Fourier representation in terms of LC time $x^\p = t +
z$ and momenta $k^\p , \prp{k}$, 
\be
  \Delta_m (x^\p, k^\p , \prp{k}) \equiv \left\{ \begin{array}{ll}
                \displaystyle  \frac{\theta (k^\p x^\p)}{i\, |k^\p |}
                e^{-i x^\p (\ssprp{k}^2 + m^2 -i \epsilon)/2 k^\p}
                \; , & k^\p \ne 0  \\ [15pt]
                \displaystyle -\frac{2\, \delta(x^\p)}{\prp{k}^2 +
                m^2 - i \epsilon} \; , & k^\p = 0 \; , \end{array}
                \right.  \label{ZMPROP} 
\ee
which displays non-uniform behavior in the longitudinal momentum $k^\p$
\cite{chang:69b}. For $k^\p \ne 0$ one has a standard Feynman--St\"uckelberg
interpretation. The \textit{zero modes} (ZMs) with  $k^\p = 0$, however,
propagate instantaneously within the null-plane $x^\p = 0$. It is hence quite
obvious that these modes will play a peculiar role in a Hamiltonian
treatment describing the time evolution \textit{off} this
null-plane \cite{maskawa:76}. While the behavior (\ref{ZMPROP}) of the
ZMs may seem strange in the first place, it is actually not entirely 
unfamiliar. Quite an analogous behavior shows up in the Coulomb gauge
propagator, 
\be
  D_{00} (x^0 , \vc{k}) = - \frac{\delta (x^0)}{\vc{k}^2 - i
  \epsilon} \; , 
\ee
which describes instantaneous `propagation' of (virtual) scalar  
gauge bosons \cite{halzen:84} resulting in the Coulomb
interaction (see below). From (\ref{ZMPROP}) we are thus led to
generally distinguish between field ZMs $\omega$ with $k^\p = 0$ and
their complement $\varphi$ which gives rise to an orthogonal
decomposition, $\phi = \varphi + \omega$, $J = \jmath + j$, $\mathscr{D}
\phi =   \mathscr{D} \varphi \, \mathscr{D} \omega$.
If we have a look at the equations of motion stemming from $S_0$ (with
sources present),
\begin{eqnarray}
  \mathsf{K}_m \, \varphi &=& (\Box + m^2) \, \varphi = \jmath \; , \\
  \mathsf{K}_m^0 \, \omega &\equiv& (- \Delta_\perp +
  m^2)  \, \omega = j \; , 
\end{eqnarray}
we note that $\omega$ obeys a constraint which is analogous to Poisson's
equation for the temporal component of the photon field (in the presence
of a static charge distribution $\rho$), $\Delta A_0 = \rho$.  Hence,
both $A_0$ and $\omega$ are `non--dynamical' fields (having no 
conjugate momenta) but are nevertheless integrated over in the path
integral. However, while the  $A_0$--integration (both in the Abelian
and non-Abelian case) is Gaussian, the $\omega$--integration in
$\phi^4$-theory, 
\be
\label{LCPI}
  Z[\jmath, j] = \int  \mathscr{D} \varphi \, \mathscr{D}
  \omega \exp i \left\{ S[\varphi , \omega] + (\varphi,
  \jmath) + (\omega , j) \right\} \; , 
\ee
is not, as the action
\be
  S[\varphi , \omega] = - \sfrac{1}{2} (\varphi, \mathsf{K}_m \,
  \varphi) - \sfrac{1}{2} (\omega, \mathsf{K}_m^0 \,
  \omega) + S_{\mathrm{int}}[\varphi , \omega]
\ee
is quartic in $\omega$. In order to have a simpler system with quadratic
action we consider the coupling to external sources. We
proceed by analogy with the Maxwell path integral in Coulomb gauge with
static source $\rho$ (or, equivalently, the vacuum expectation value of
the Abelian Wilson loop). The latter is proportional to the integral 
\be
  \int \mathscr{D}A_0 \, \exp i
  S_{\rho} [A_0] \; \sim \; \exp \left\{ -i T E[\rho] \right\} \; , 
\ee 
with quadratic action $S_{\rho} = - \sfrac{1}{2} (A_0 , \Delta A_0) +
(\rho, A_0)$  and yields  the Coulomb potential (plus an irrelevant
divergent self-energy that is independent of any length scales
associated with the source),
\be
  E[\rho] = -\frac{(\rho, D_{00} \, \rho)}{2T} = \frac{1}{2} \int
  \frac{d^{d-1}k}{(2\pi)^{d-1}} \, \frac {|\rho(\vc{k})|^2}{\vc{k}^2} =
  \frac{1}{4\pi r} + const \; ,
\ee  
where the last identity holds for two unit point charges at distance $r$
in $d=4$. Invoking covariance, the LC path integral with ZM
source $j$ should have the asymptotic behavior,
\be
\label{LCPI_SOURCE}
  \int \mathscr{D}\omega \, \exp i S_{j} [\omega]  \;
  \sim \; \exp \left\{-i L P^\p [j] \right\} \; ,  
\ee
where we assume a quadratic action $S_{j}  = - \sfrac{1}{2} (\omega,
\mathsf{K}_m^0 \, \omega) +  (\omega, j)$. Performing the Gaussian
integral (\ref{LCPI_SOURCE}) results in a source-generated longitudinal
momentum proportional to a transverse Yukawa potential, 
\be 
  P^\p [j] = \frac{(j, \Delta_m^0 \, j)}{2L} = - \frac{1}{4} \int
  \frac{dk^\m}{2\pi} \int \frac{d^{d-2}k_\perp}{(2\pi)^{d-2}}
  \, \frac{|j(k^\m , \prp{k})|^2}{\prp{k}^2 + m^2} = - \frac{1}{2\pi} \,
  \delta(z^\p) \, K_0 (m z_\perp) + const \; .
\ee 
Again, the last equality holds for two point sources of unit strength at
distance $(z^\p , \prp{z})$ in $d=4$. Furthermore, in the first
equality, we have introduced the ZM propagator
\be
   \Delta_m^0 (x-y) \equiv - \int \frac{d^d k}{(2\pi)^d} \, \frac{e^{i k
   \cdot (x-y)}}{\prp{k}^2 + m^2} \equiv
   \DashLine(8,3)(50,3){4} \Text(10,-5)[]{$x$} \Text(48,-7)[]{$y$}
   \hspace{2.2cm} \; ,
\ee
which will become important in a moment. 

\section{ZERO MODES}

Let us go back to the more realistic situation of $\phi^4$-theory. There
are two methods by which the zero modes can be integrated
out approximately, perturbation theory and the saddle point
approximation. 

\subsection{Perturbation theory}

In this case one expects that the integration over $\omega$ corresponds
to the limiting case of a Wilsonian momentum--shell integration over
momenta from the interval $0 \le k^\p \le \delta$. This should give rise
to an RG interpretation, which will be developed elsewhere. Here, we
content ourselves with displaying the lowest-order  effective interactions
induced by the $\omega$--integration. They typically include the ZM
propagator $\Delta_m^0$, either within vacuum bubbles,  

\clearpage

\begin{figure}[h]
  \hspace{2cm}  \includegraphics[scale=.8]{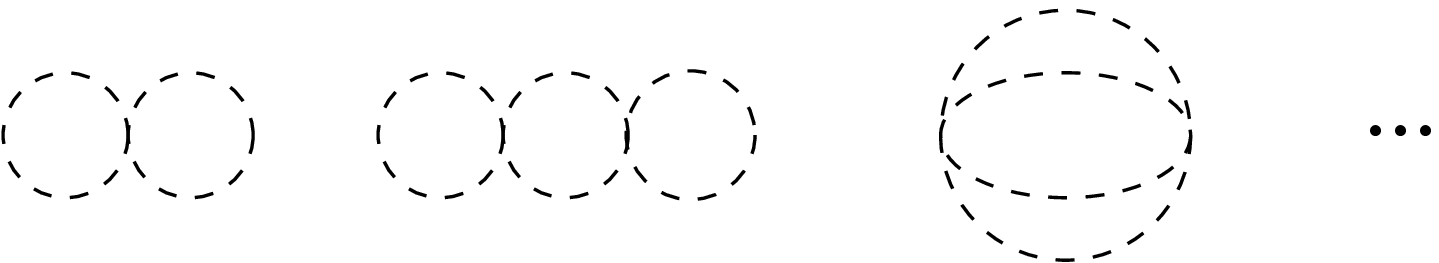}
\end{figure}

\noindent
or in terms of new (nonlocal) $\varphi$--vertices, 
\begin{figure}[h]
  \hspace{1cm} \includegraphics[scale=.7]{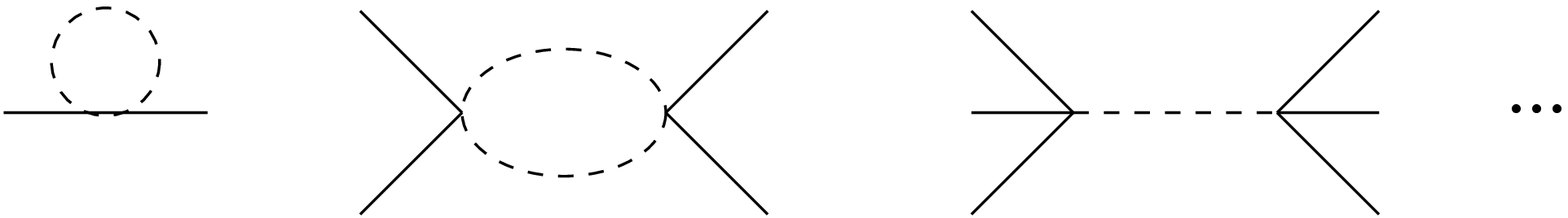}
\end{figure}

\noindent
On the other hand, solving the constraint equation, 
\be
\label{PHI4CONSTRAINT}
  \omega =  \frac{\lambda}{3!} \frac{1}{2L} \intl dx^\m \,
  \Delta_m^0  * (\varphi +\omega)^3   \quad \quad (L \to \infty) \; , 
\ee
in perturbation theory \cite{mccartor:92} 
corresponds to ZM tree level as ZM loops are neglected
\cite{taniguchi:01}. In other words, the ZM sector is treated
classically, i.e. to order $\hbar^0$. For instance, if one considers the
following 2-by-2 scattering amplitude,

\begin{figure}[ht] 
\begin{center} 
  \includegraphics[scale=.65]{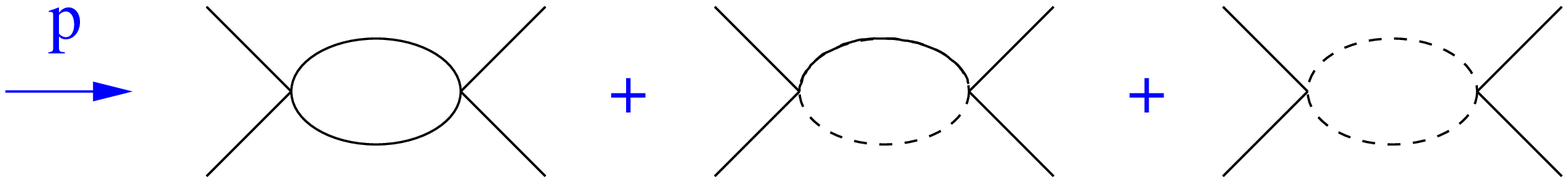}
\end{center}
\end{figure}

\noindent 
one finds that only the first two diagrams are obtained by solving the
constraint. They both vanish for $p^\p \to 0$ \cite{taniguchi:01}. Only
the last (ZM loop) diagram survives this limit. Unfortunately, this
argument does not constitute a proof that ZMs cannot be neglected. In
standard (noncovariant) LC perturbation theory the two time-orderings
representing the first diagram have the correct covariant limit for
$p^\p \to 0$ \cite{harindranath:02}. It would be intersting to study the
critical behavior of $\phi^4$--theory using the new Feynman rules and
compare with the results of \cite{SGW:02} which were obtained by solving
the constraint.

\subsection{Semiclassical approximation} 

Alternatively, one may integrate  out $\omega$ by expanding the action
$S[\varphi, \omega]$ around the saddle point $\phi_0 = \omega_0 +
\varphi_0 = \omega_0$, where we assume that $\omega_0$ is a
\textit{constant} solution of the classical equation of motion
(\ref{PHI4CONSTRAINT}). Again, the classical approximation neglects
quantum fluctuations around $\omega_0$, and approximates the VPA by
\be
\label{CLASSAPPROX}  
  Z[0] \simeq \int \mathscr{D} \varphi \, \exp i S[\varphi ,
  \omega_0] = \int \mathscr{D} \varphi \int  \mathscr{D}\omega \, \delta
  (\omega - \omega_0) \exp i S[\varphi , \omega] \; .
\ee
This is exactly the path integral which is obtained  via the
Dirac-Bergmann algorithm for constrained systems \cite{heinzl:91}. If
one wants to go beyond (\ref{CLASSAPPROX}), one has to calculate
the fluctuation determinant. For constant background $\omega_0$ this
should yield the effective potential.

\section{THE EFFECTIVE POTENTIAL}

Indeed, the LC path integral (\ref{LCPI}) yields the standard formula 
$V_{\mathrm{eff}} \equiv V_0 + V_1$,  with $V_0 [\omega_0]$ being the
classical potential and $V_1 [\omega_0]$ the one-loop effective
potential,  
\be
  V_1 \equiv - \frac{i}{2} \, \Omega_d^{-1} \, \Tr \, \log \, (\Delta_m /
  \Delta_{\mu}) \; . 
\ee
Here, we have introduced the notation $\Omega_d \equiv \mbox{vol}
(\mathbb{R}^d)$ and $\mu^2 \equiv m^2 + \lambda \, \omega_0^2 / 2$. 
\enlargethispage{\baselineskip}

\subsection{Calculation \textit{with} ZMs}

Expanding the log yields the 1PI diagrams

\begin{minipage}[c]{1.5cm} 
$\Omega_d V_1 =$
\end{minipage} 
\begin{minipage}[c]{12cm}  
\begin{picture}(2,12) 
\put(110,22){$\left\{ \rule[-.9cm]{0pt}{1.5cm} \hspace{6cm} \right\}$} 
\end{picture} 
  \includegraphics[scale=0.45]{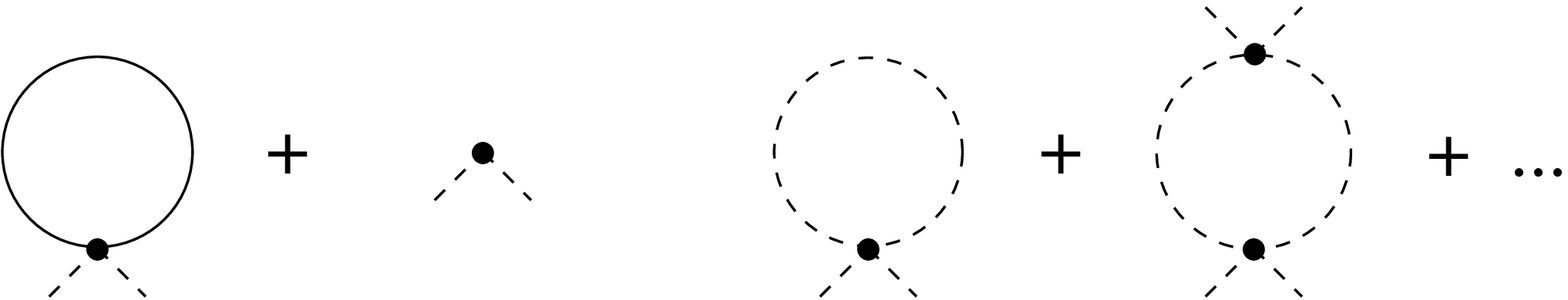}
\end{minipage}

\vspace{.5cm} 
\noindent 
where the first diagram is the standard covariant tadpole $T_m =
\Omega_d^{-1} \Tr \, \Delta_m$ of mass $m$, while the remaining ones are
exclusively given by ZM loops. Each vertex attached to external
$\omega$-legs represents the factor $\lambda \, \omega_0^2 / 2$. The
calculation of the tadpole $T_m$ is done in standard dimensional
regularisation (dimReg).  The ZM bubbles are evaluated by means of the
formula, 
\be 
  \int \frac{dk^\m}{2\pi i} \, (k^\p k^\m - M^2 + i \epsilon)^{-p}
  = \frac{(-1)^{p}}{p-1}
  \, \frac{\delta(k^\p)}{(M^2)^{p-1}} \; , \quad p \ne 1 \; , 
\ee
which has first been used by Yan for $p=2$ \cite{yan:73b}. The
$k_\perp$-integrations are done via dimReg. Altogether, 
we reproduce the standard dimReg result for the effective potential.  

\subsection{Calculation \textit{without} ZMs}  

The idea is to relate the effective potential $V_{\mathrm{eff}}$ to the
tadpole $T_{\mu}$ via differentiation with respect to the ZM-induced mass
$\mu^2$, 
\be 
  \pad{V_1}{\mu^2} = \frac{i}{2} \int \frac{d^d k}{(2\pi)^d} \, 
  \frac{1}{k^2 - \mu^2 + i \epsilon} \equiv \frac{i}{2}
  \, T_{\mu} \; ,
\ee 
so  that we are left with a single divergent tadpole
integral. Intuitively, the tadpole can be viewed as the (infinite)
volume of the mass--shell $k^2 = \mu^2$, see Fig.~1.

\begin{figure}[ht]
\begin{center}
  \includegraphics[scale=.35]{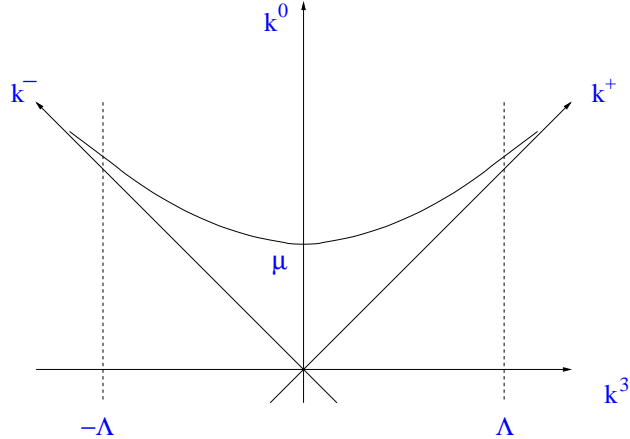} 
    \caption{Mass-shell and noncovariant cutoffs ($d=2$).}   
\end{center}  
\end{figure} 

\noindent 
It is obvious that a noncovariant cutoff $-\Lambda < |\vc{k}| <
\Lambda$ renders the tadpole  finite and gets translated in a cutoff for
both large and small $k^\p$ \cite{harindranath:88}, $\mu^2 / \Lambda  <
k^\p < \Lambda$, which eliminates any ZM contribution, and a transverse
momentum cutoff \cite{dietmaier:89}, $\prp{k}^2 < \Lambda^2 x (1-x) -
\mu^2$ with $x \equiv k^\p /\Lambda$. Employing this regularization, one
obtains the standard cutoff result for $V_{\mathrm{eff}}$ in $d=2$ and
$d=4$ \cite{heinzl:02}.
 
\section{SUMMARY AND OUTLOOK}

Resuming it seems fair to say that the physics of LC ZMs still remains
somewhat elusive. Nevertheless, the LC path integral provides (new)
intuition. A trivial vacuum may be seen as resulting from the
subtraction of a `cosmological constant'.  After normal-ordering (which
eliminates $\varphi$-loops), this constant must be entirely due to ZM
loops. These, however, are also zero if
one sticks to ordinary constrained dynamics (or, equivalently, the
classical `approximation') in the ZM sector as has been common usage
before. Effective ZM loops arise if one integrates out the ZM $\omega$
in the LC path integral in perturbation theory. A semiclassical
approximation yields the standard effective potential. Details of the
latter calculation show that summing the ZM contributions amounts to
tuning the mass-dependent cutoff ($m^2 \to \mu^2$), which seems to
suggest some  RG interpretation. Of course, it would be desirable to
make contact with a Hamiltonian formulation. We believe that the latter
can only be defined with an intrinsic cutoff $k^\p > \delta$, making the
vacuum trivial. While this removes ZMs, covariance  
presumably requires the presence of new interactions with $k^\p =
0$. The most interesting quantities one wants to know are the LC wave
functions. Recently it has been shown that these can be extracted from the
large-LC-time asymptotics of the LC Schr\"odinger functional
\cite{petrov:03}. The study of this functional is interesting by itself
as it is one of the keys to Hamiltonian renormalization.  

\section*{ACKNOWLEDGEMENTS}

Part of this work has been done together with P. Grang{\'e} whose collaboration
is gratefully acknowledged. The author thanks Universit{\'e} Montpellier
II for support and the workshop organizers (in particular S.~Dalley)
for creating such an enjoyable workshop atmosphere. Partial funding was
provided by DFG under contract Wi-777/5-1.

\end{document}